\documentclass{emulateapj}
\slugcomment{}

\shorttitle{FRBs from Binary WD Mergers}
\shortauthors{K. Kashiyama, K. Ioka, and P. M$\rm \acute{e}$sz${\rm \acute{a}}$ros}

\begin{document}

\title{Cosmological Fast Radio Bursts from Binary White Dwarf Mergers}

\author{Kazumi Kashiyama\altaffilmark{1}}
\email{kzk15@psu.edu}
\author{Kunihito Ioka\altaffilmark{2}}
\email{kunihito.ioka@kek.jp}
\author{Peter M$\rm \acute{e}$sz${\rm \acute{a}}$ros\altaffilmark{1}}
\email{nnp@psu.edu}

\altaffiltext{1}{Department of Astronomy \& Astrophysics; Department of Physics; Center for Particle \& Gravitational Astrophysics; Pennsylvania State University, University Park, PA 16802}
\altaffiltext{2}{Theory Center, Institute of Particle and Nuclear Studies, KEK; Department of Particle and Nuclear Physics, the Graduate University for Advanced Studies (Sokendai), Tsukuba 305-0801, Japan}

\begin{abstract}
Recently, Thornton et al. reported the detection of four fast radio bursts (FRBs).  
The dispersion measures indicate that the sources of these FRBs are at cosmological distance. 
Given the large full sky event rate $\sim 10^4 \ \rm sky^{-1} day^{-1}$, the FRBs are a promising target for multi-messenger astronomy.
Here we propose double degenerate, binary white-dwarf~(WD) mergers as the source of FRBs, 
which are produced by coherent emission from the polar region of a rapidly rotating, magnetized massive WD formed after the merger. 
The basic characteristics of the FRBs, such as the energetics, emission duration and event rate, can be consistently explained in this scenario.
As a result, we predict that some FRBs can accompany type Ia supernovae~(SNe Ia) or X-ray debris disks. 
Simultaneous detection could test our scenario and probe the progenitors of SNe Ia, and moreover would provide a novel constraint on the cosmological parameters. 
We strongly encourage future SN and X-ray surveys that follow up FRBs. 

\end{abstract}


\section{Introduction}
\cite{Thornton_et_al_2013} recently reported the detection of four fast radio bursts (FRBs),
discovered in the high Galactic latitude region of the High Time Resolution Universe survey~\citep{Keith_et_al_2010} 
using the 64-m Parks radio telescope at $\sim 1.5 \ \rm GHz$.  
The characteristics of FRBs can be summarized as follows~\citep{Thornton_et_al_2013};
\begin{itemize}
\item The dispersion measures are ${\rm DM} \gtrsim 500\mbox{-}1000 \ \rm cm^{-3} \ pc$, which implies that the 
sources are at cosmological distances with redshifts $z \gtrsim 0.5\mbox{-}1$. 
\item The typical flux is $S_{\nu}~\sim \ \rm Jy$.~\footnote{The so-called Lorimer burst (FRB 010724) is 
$\sim~10\mbox{-}100$ times brighter than the four FRBs~\citep{Lorimer_et_al_2007}.}
The total emitted energy can be calculated as $\sim 10^{38\mbox{-}40} \ \rm erg$, if the distance is cosmological and the emission is nearly isotropic.  
\item The observed durations of the bursts are $\delta t = 5.6 \ \rm ms$, $< 4.3 \ \rm ms$, $< 1.4 \ \rm ms$, and $< 1.1 \ \rm ms$, 
which indicate that the emission regions are relatively compact, $c \delta t (1+z)^{-1} \lesssim 1500 \ (1+z)^{-1} \ \rm km$, if they are non-relativistic. 
\item The observed rate is $\sim~(1\pm 0.5)~\times~10^4 \ \rm sky^{-1} \ day^{-1}$. 
This can be translated into $R_{\rm FRB} \sim 10^{-3}  \ \rm yr^{-1} \ galaxy^{-1}$ for late-type galaxies, 
which is roughly $10 \ \%$ of that of core-collapse supernovae~(CCSNe), $R_{\rm CCSN} \sim 10^{-2}~\ \rm yr^{-1} \ galaxy^{-1}$.     
\item No repeated bursts have been reported thus far, so they may be one-time-only events.   
\item No counterparts in other messengers have been reported. 
\end{itemize}

Several possibilities for the progenitors of FRBs have been proposed, e.g., magnetar giant flares~\citep{Popov_Postnov_2007, Thornton_et_al_2013}, 
SN explosion in a neutron-star~(NS) binary system~\citep{Egorov_Postnov_2009}, 
accretion-induced collapse of hypermassive NSs into black holes~\citep{Falcke_Rezzolla_2013}, 
binary NS mergers~\citep{Totani_2013}, and evaporation of primordial black holes~\citep{Keane_et_al_2012}. 
Multi-messenger follow-up observations are crucial to discriminate between the above models. 

In this Letter we propose binary white dwarf~(WD) mergers as a promising candidate for origin of the FRBs.
We consider the coherent emission in the polar region of rapidly rotating, magnetized massive WDs formed just after the merger. 
Observationally, up to 10 \% of WDs have been confirmed to be magnetized, 
and $\sim$ 1 \% of them have a strong magnetic field of $B \sim 10^{8\mbox{-}9} \ \rm G$~\citep[e.g.,][]{Kawka_et_al_2007, Kepler_et_la_2013}. 
Since such strongly magnetized WDs are typically more massive than the average of the total population~\citep{Liebert_et_al_2003}, 
they are considered to be formed by binary WD mergers. 

Our model can reproduce the main observed characteristics of FRBs, such as the energetics, emission duration, and event rate. 
As an offshoot, we predict that some FRBs can accompany type Ia supernova~(SN Ia) events. 
If the redshift or the distance to individual FRBs can be determined by identifying their host galaxies or SN Ia counterparts, 
this could open a new possibility for exploring the cosmological parameters and cosmic reionization history by using the dispersion measure of FRBs. 

\section{Energetics, Timescale and Rate}
\begin{figure}
\includegraphics[width=90mm]{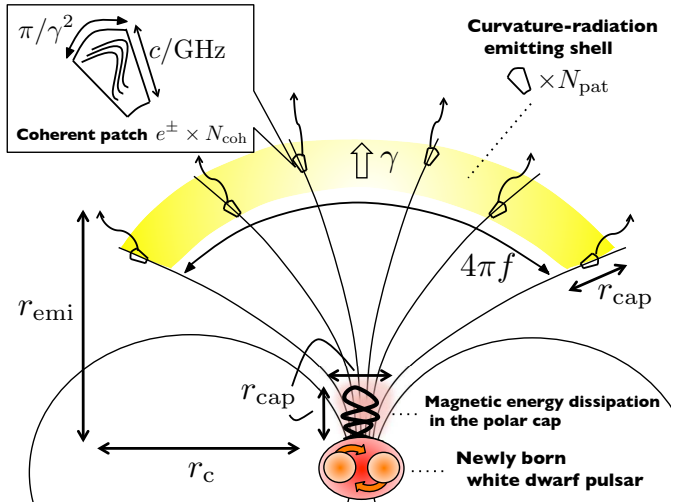}
\caption{Schematic picture of binary white dwarf merger model for fast radio bursts.  
}
\label{mdot}
\end{figure}

Double degenerate binary-WD mergers have been investigated in various contexts. 
They are particularly interesting as the progenitors of  SNe Ia~\citep[see, e.g.,][]{Hillebrandt_Niemeyer_2000}. 
For non-explosion cases, the remnant is expected to be a rapidly rotating, magnetized massive WD. 
The delayed accretion from the surroundings may induce a collapse into a NS or a magnetar, which may trigger a short gamma-ray bursts~\citep[GRBs;][]{Levan_et_al_2006, Metzger_et_al_2008}. 
Even if the above explosive phenomena do not occur, the remnant WDs, so-called WD pulsars~\citep[e.g.,][]{Terada_et_al_2008}, 
can be significant sources of GeV-TeV electron-positron cosmic rays during the spin-down period~\citep{Kashiyama_et_al_2011}, 
which may explain the electron-positron excesses observed by 
PAMELA~\citep{Adriani_et_al_2009}, AMS-02~\citep{Aguilar_et_al_2013}, Fermi~\citep{Ackermann_et_al_2010}, HESS~\citep{Aharonian_et_al_2008}, and MAGIC~\citep{Borla_2011}. 
Finally, gravitational waves from binary WDs in the in-spiral and merger phase should be detectable by {\it LISA} and DECIGO~\citep{Nelemans_et_al_2001,Farmer_Phinney_2003}. 
Hence, binary WD mergers may be one of the most interesting targets in the coming era of multi-messenger astronomy 
, e.g., with specially designed facilities such as AMON~\citep{Smith_et_al_2013}.
In the following, we show that they are also plausible candidates for cosmological FRBs. 

First, we show that a binary WD merger can provide the right energy budget for a FRB. 
We consider a rapidly rotating massive WD formed just after a WD-WD merger. 
The rotation velocity can be as high as the mass-shedding limit, $v \approx (GM/r)^{1/2} \sim 5.7 \times 10^8 \ r_{8.7}{}^{-1/2} \ \rm cm \ s^{-1}$, 
and the angular frequency can be estimated as $\Omega \approx v/r \sim 1.1 \ r_{8.7}{}^{-3/2} \ \rm s^{-1}$ (hereafter we use $Q_x=Q/10^x$ in cgs units).
In numerical simulations, the rotation velocities of the merged WDs are a factor of $3\mbox{-}4$ smaller than the above value~\citep[e.g.,][]{Loren_et_al_2009}.  
We set the nominal mass of the merged WD as $M \sim 1.2 M_{\odot}$, using the average value $\sim 0.6 M_{\odot}$ for the observed WDs~\citep{Falcon_et_al_2010}.   
Also, as the typical radius, we use $r = 10^{8.7} \ \rm cm$, which corresponds to a cold WD with a mass of $\sim M_{\odot}$.
Given the possible emission mechanism (see below), the kinetic energy is transferred first into magnetic fields and into electron acceleration, and finally into electromagnetic radiation.  
If the merged object rotates differentially, the magnetic field may be amplified by magnetorotational instabilities or dynamo processes~\citep[e.g.,][]{Shibata_et_al_2011,Ji_et_al_2013}.  
Indeed, numerical simulations show that the magnetic fields are amplified up to $B \sim 10^{7\mbox{-}9} \ \rm G$ after the merger~\citep{Kulebi_et_al_2013,Ji_et_al_2013}. 
Hereafter, we take $B \sim 10^{9} \ \rm G$ as a fiducial surface magnetic-field strength.  
The total magnetic-field energy outside the WD can then be estimated as 
\begin{equation}\label{eq:ene}
E_{\rm B} \approx (B^2/8\pi) \times (4 \pi r^3/3) \sim 2.1 \times 10^{43} \ B_9{}^2 r_{8.7}{}^3 \ \rm erg. 
\end{equation}
Thus, a fraction $\sim 5\times 10^{-3} \ f B_9{}^2 r_{8.7}{}^3$ of the energy is enough for the total emitted energy in typical FRBs, 
where the pre-factor $f$ is the beaming of the radio emission. 
Note that the magnetic fields could be amplified further, up to a maximum value at which the local gravitational binding can support them, i.e., 
$(B_{\rm max}^2/8\pi) \times (4 \pi r^3/3) \lesssim GM^2/r \sim 7.7 \times 10^{50} \ \rm erg$.  
Thus, the possible transient maximum magnetic-field strength is $B_{\rm max} \sim 6.1 \times 10^{12} \ \rm G$, 
which is comparable to the surface field of ordinary NS pulsars.  

Next, we show how the observed duration of the FRBs can be reproduced in our scenario. 
The inferred sizes of the emission regions, 
\begin{equation}
c \delta t (1+z)^{-1} \lesssim 1.5 \times 10^8 \ (1+z)^{-1} \ \rm cm, 
\end{equation}
are relatively small compared with our fiducial WD radius, $r \sim 10^{8.7} \ \rm cm$. 
Thus, the emission has to originate from a portion of the merged WD. 
Interestingly, the transverse size of the polar-cap region, $r_{\rm cap} \approx r (r\Omega/c)^{1/2} 
\sim 6.9 \times 10^7 \ r_{8.7}{}^{3/2} \Omega_{0}{}^{1/2}\ \rm cm$ for a break-up angular frequency 
$\Omega_0\sim 1~{\rm s}^{-1}$, is in a reasonable range, and the resultant light crossing time, 
\begin{equation}\label{eq:time}
r_{\rm cap}/c \sim 2.3 \ r_{8.7}{}^{3/2} \Omega_{0}{}^{1/2} \ \rm ms, 
\end{equation}
is shorter than the observed (dispersed) duration. 
The region above the polar-cap is also a plausible site for production of coherent radio emission, as we argue below.  
An electric-field potential as large as that of ordinary NS pulsars can be produced via the unipolar induction mechanism~\citep{Goldreich_Julian_1969,Ruderman_Sutherland_1975};
\begin{equation}\label{eq:ele}
\Phi_{\rm max} \approx B\Omega^2r^3/2c^2 \sim 2.5 \times 10^{16} \ B_9 \Omega_0{}^2 r_{8.7}{}^3 \ \rm Volt,
\end{equation}
and pair-production avalanches can be triggered in the strong magnetic field~\citep{Kashiyama_et_al_2011}. 

We note that the spin-down luminosity of the magnetized WD, $L_{\rm WD} \sim 1.7 \times 10^{38} \ B_9{}^2 r_{8.7}{}^6 \Omega_0{}^4 \rm erg s^{-1}$, 
is significantly smaller than the FRB isotropic luminosity $L_{\rm FRB} \sim 10^{43\mbox{-}44} \ \rm erg \ s^{-1}$. 
\footnote{In the case of newborn NSs, the spin-down luminosity can be consistent with the FRB luminosity~\citep{Falcke_Rezzolla_2013,Totani_2013}.}. 
Thus, our scenario requires that the magnetic field energy is transiently released in the polar-cap region via processes such as reconnection.  
Such processes are in fact generally present in strongly magnetized astrophysical objects~\citep{Thompson_Duncan_1995}.  
For example, soft gamma-ray repeater or magnetar flares would be rapid transient magnetic dissipation events
(as in our case, the spin-down luminosity of magnetars is also much smaller than the luminosity of such flares).
Also, solar flares are explained by transient magnetic dissipation events in localized regions on the surface.  
They are triggered by the emergence of magnetic fields from the stellar interior, 
which can have orders-of-magnitude higher values than the average field value~\citep[e.g.,][]{Shibata_Magara_2011}.  
In our case, if the inner magnetic field of, e.g., $B_{\rm in} \sim 10^{10} \ \rm G$, is convectively transported up into a polar cap region of volume  
$V_{\rm cap} \approx \pi r_{\rm cap}{}^3 \sim 1.0 \times 10^{24} \ r_{8.7}{}^{9/2} \Omega_{0}{}^{3/2}  \ \rm cm^3$,  
the corresponding energy budget becomes $(B_{\rm in}{}^2/8\pi) \times V_{\rm cap} \sim 4.0 \times 10^{42} \ B_{\rm in, 10}{}^2  r_{8.7}{}^{9/2} \Omega_{0}{}^{3/2} \ \rm erg$, which is enough for a FRB.
The dissipation timescale would be given by Equation (\ref{eq:time}), which determines the injection timescale of electron bunches into the polar region (see below), 
reproducing the observed duration.

Next, let us consider the event rate. 
Reasonable estimates of the Galactic WD-WD merger rate are in the range \citep{Badenes_Moaz_2012}, 
\begin{equation}
R_{\rm WD} \sim 10^{-2}\mbox{-}10^{-3} \  \rm yr^{-1} galaxy^{-1}, 
\end{equation}
though the uncertainties are large, e.g., in the initial mass function, the distribution of the initial separation, and the evolution of the binary during periods of non-conservative mass transfer.
The angular size of the polar cap is relatively small, $\theta_{\rm cap} \approx (r\Omega/c)^{1/2} \sim 0.14 \ r_{8.7}{}^{1/2} \Omega_{0}{}^{1/2}$, 
which would imply a surface beaming factor $f_{\rm cap} \approx \theta_{\rm cap}{}^2/4 \sim 4.9 \times 10^{-3} \ r_{8.7} \Omega_{0}$. 
However, the beaming factor at the emission region is likely larger, $f \gtrsim 0.1$. 
As shown below, the emission radius has to be an order-of-magnitude larger than the WD radius, e.g., $r_{\rm emi} \sim 10^{10} \ \rm cm$, 
for the GHz radio emission to escape the emitter, where a larger beaming factor is expected. 
The other factor is the evolution of the WD-WD merger rate, 
which can be significantly larger at redshifts $z \sim 1$ than the local rate~\citep[e.g.,][]{Schneider_et_al_2001}.  
Combining all of the above, the anticipated detection rate can be considered consistent with the observed FRB rate.   

\section{Emission Mechanism}\label{sec:emi}
The FRBs are likely coherent emission as indicated by the high brightness temperatures of $\sim 2.8 \times 10^{41} \ L_{\rm FRB, 43} r_{\rm emi, 10}{}^{-2} \ \rm K$, where $r_{\rm emi}$ is the emission radius.
The mechanism would be similar to the radio pulses from ordinary NS pulsars, including giant pulses, whose mechanism is still highly uncertain~\citep[e.g.,][]{Lyubarsky_2008}. 
Below we examine one of the promising models for coherent emission, the coherent curvature emission model~\citep{Ruderman_Sutherland_1975,Bushauer_Benford_1976}.  

Let us consider electron bunches forming a shell with an average thickness $\delta r_{\rm emi}$, a transverse size $r_{\rm emi}$, a Lorentz factor $\gamma$, 
and an electron density $n_{\rm e}$ in the observer frame, coasting along a magnetic field with a typical curvature radius $r_{\rm c}$. 
The width of the emitting shell is determined by the duration of the central engine activity, i.e., $\delta r_{\rm emi} \approx c \delta t$, 
which would be roughly constant in the observer frame as long as $r_{\rm emi} < \delta r_{\rm emi} \gamma^2$ like GRB fireballs~\citep{Rees_Meszaros_1992}. 

The frequency of the curvature radiation is $\nu_{\rm c} \approx \gamma{}^3 (3c/4\pi r_{\rm c}) 
\sim 0.72 \ \gamma_{3}{}^3 r_{\rm c, 10}{}^{-1} \ \rm GHz$. Thus,  in order to explain the 
observed frequency of $\nu_{\rm FRB} \sim$ GHz, one needs
\begin{equation}\label{eq:fre}
\gamma \gtrsim 1100 \ \nu_{\rm FRB, 9}{}^{1/3} r_{c, 10}{}^{1/3}. 
\end{equation}

The total luminosity of the curvature radiation from the emitting shell can be 
expressed as
\begin{equation}
L_{\rm tot} \approx  (P_{\rm c} N_{\rm coh}{}^2) \times N_{\rm pat}.  
\end{equation}
Here $P_{\rm c} \sim 2\gamma^4 e^2 c/ 3r_{\rm c}^2 \sim 4.6 \times 10^{-17} \ \gamma_3{}^4 r_{c, 10}{}^{-2} \ \rm erg \ s^{-1}$ is the power radiated by each electron, 
and  $N_{\rm coh} \approx n_{\rm e} \times V_{\rm coh} \sim 10^{23} n_{\rm e, 7} V_{\rm coh, 16}$ denotes the maximum number of electrons in a patch  producing coherent emission. 
The volume of such coherent patches is given by $V_{\rm coh} \approx (4/\gamma^2)r_{\rm emi}{}^2 \times (c/\nu_{\rm c}) \sim 1.7 \times 10^{16} \ 
\gamma_3{}^{-5} r_{\rm c, 10}  r_{\rm emi, 10}{}^2  \ \rm cm^3$.  
We note that $c/\nu_{\rm c}$ corresponds to the radial width of the patch, and only the electrons within a solid angle of $\approx 4/\gamma^2$ can be causally connected due to relativistic beaming. 
The number of such coherent patches in the emitting shell can be estimated as 
$N_{\rm pat} \approx V_{\rm emi}/V_{\rm coh} \sim 5.2 \times 10^{12}  \ f  \gamma_3{}^{5} r_{\rm c, 10}{}^{-1} \delta r_{\rm emi, 7.8}$. 
Here, $V_{\rm emi} \approx 4 \pi f r_{\rm emi}{}^2 \times \delta r_{\rm emi} \sim 8.6 \times 10^{28} \ f r_{\rm emi,10}{}^2 \delta r_{\rm emi, 7.8}  \ \rm cm^3$ 
is the total volume of the emitting shell with the beaming factor $f$.  
Thus, we have
\begin{equation}
L_{\rm tot} \sim 4.0 \times 10^{42} \ f n_{\rm e,7}{}^2 \gamma_3{}^{-1} r_{\rm c,10}{}^{-1} r_{\rm emi,10}{}^4 \delta r_{\rm emi, 7.8} \ \rm erg \ s^{-1}. 
\end{equation}
To explain the observed FRB luminosities, $L_{\rm FRB} \sim 10^{41\mbox{-}43} \ f \ \rm erg \ s^{-1}$, one needs
\begin{eqnarray}\label{eq:lum}
n_{\rm e} &\sim& 1.6 \times 10^7 \ \rm cm^{-3}  \nonumber \\ &\times&  L_{\rm FRB,43}{}^{1/2} \gamma_3{}^{1/2} r_{\rm c,10}{}^{1/2} r_{\rm emi,10}{}^{-2} \delta r_{\rm emi, 7.8}{}^{-1/2}.  
\end{eqnarray}

Importantly, only the radio emission at frequencies above the plasma frequency can escape the emitter without converting into plasma waves, i.e.,  
\begin{equation}\label{eq:plas}
\nu_{\rm c} \gtrsim \nu_{\rm p} \approx \frac{\gamma}{2\pi} \left( \frac{4 \pi n'_{\rm e} e^2}{m_{\rm e}} \right)^{1/2}. 
\end{equation}
Here $n'_{\rm e} = n_{\rm e}/\gamma$ is the electron number density in the comoving frame. 
To satisfy Equation (\ref{eq:plas}), one needs 
\begin{equation}\label{eq:esc}
n_{\rm e} \lesssim 0.64 \times 10^7 \ \gamma_{3}{}^5 r_{\rm c, 10}{}^{-2} \ \rm cm^{-3}. 
\end{equation}

In addition, the induced Compton scattering can prevent coherent emission from being observed~\citep[e.g.,][]{Melrose_1971,Wilson_Rees_1978}. 
One can neglect this effect if the relevant timescale for the scattering is longer than the dynamical timescale of the emitter~\citep[Equations (42) and (47) of][]{Lyubarsky_2008b};  
\begin{equation}\label{eq:induced_comp_1}
\frac{2m_{\rm e} \gamma_T \nu'{}^2}{3c\sigma_{\rm T} n'_{\rm e} I'} > \frac{r_{\rm emi}}{c\gamma}. 
\end{equation}
Here, $I'$, $\nu'$, and $\gamma_T$ are the flux and the frequency of the photon bunch, and the thermal Lorentz factor of the electrons in the comoving frame, respectively. 
For coherent curvature radiation, $I'$ can be estimated from  $P_{\rm c} N_{\rm coh}{}^2 = (4/\gamma^2)r_{\rm emi}{}^2 \times 4 W' c \gamma^2$, 
with the radiation density, $W' \approx 4\pi I' \nu'/c$, and one can set $\nu' = \nu_{\rm c}/\gamma$, $\gamma_T \approx 1$.  
Then, the condition (\ref{eq:induced_comp_1}) can be written as 
\begin{equation}\label{eq:induced_comp_2}
\gamma \gtrsim 2000 \ r_{\rm c, 10}{}^{3/14} r_{\rm emi, 10}{}^{3/14} n_{\rm e, 7}{}^{3/14}. 
\end{equation}

From Equations (\ref{eq:fre}), (\ref{eq:lum}), (\ref{eq:esc}), and (\ref{eq:induced_comp_2}), one can explain the observed characteristics of the FRBs by coherent curvature radiation from emitting shells with 
$\gamma \gtrsim 10^3$, $r_{\rm emi} \approx r_{\rm c} \gtrsim 10^{10} \ \rm cm$, and $n_{\rm e} \lesssim 10^7 \ \rm cm^{-3}$. 
In this case, the dispersion measure in the emitter $\sim n_{\rm e} r_{\rm emi} \sim 0.03 \ n_{\rm e, 7} r_{\rm emi, 10}{}^{-1} \ \rm pc \ cm^{-3}$ is negligible compared to cosmological dispersion. 

In our model, first, the magnetic fields in the merged WD are amplified in a dynamical timescale $\sim \Omega^{-1} \sim 0.91 \  r_{8.7}{}^{3/2} \ \rm s$ after the merger. 
With a comparable timescale, interior magnetic fields emerge on the surface by convection. 
In the polar-cap region, they are twisted by differential rotations of the surface or magnetic instabilities, which trigger dissipation processes such as reconnection, and inject electron bunches 
with a timescale of $\approx r_{\rm cap}/c \sim 2.3 \ r_{8.7}{}^{3/2} \Omega_{0}{}^{1/2} \ \rm ms$ (Equation (\ref{eq:time})).   
The electron bunches are accelerated in the polar electric fields or by the large internal energy as GRB fireballs~\citep{Rees_Meszaros_1992}. 
The electron-positron numbers are multiplied by photon-pair and/or magnetic-pair production as in NS pulsars~\citep{Kashiyama_et_al_2011}. 
These electrons coast along the open magnetic fields, and emit coherent curvature radiation~(see Figure 1).  

First, we note that $r_{\rm emi} \approx r_{\rm c} \sim 10^{10} \ \rm cm$ is close to the light cylinder of newly born WD pulsars, $r_{\rm lc} \approx 3 \times 10^{10} \ r_{8.7}{}^{3/2} \ \rm cm$.  
At these radii, the polar magnetic field lines become open, which is a plausible site for curvature radiation.  
This also means that the beaming factor of the emitting shell is relatively large, e.g., $f \gtrsim 0.1$. 

At the emission radius, the acceleration rate of an electron can be estimated as $\approx e B_{\rm emi} r_{\rm emi} \Omega \sim 5.3 \times 10^5 \ B_{\rm emi, 5} r_{\rm emi, 10} \Omega_0\ \rm erg \ s^{-1}$. 
Here, $B_{\rm emi}$ is the magnetic-field strength at the emission radius, which can be estimated as 
$B_{\rm emi} \sim B \times (r_{\rm emi}/r)^{-3} \sim 1.3 \times 10^5 \ B_9 r_{\rm emi, 10}^{-3} r_{8.7}{}^{3} \ \rm G$ for the dipole case.  
On the other hand, the energy loss rate of the electron is $\approx P_{\rm c} N_{\rm coh} \sim 7.8 \times 10^6 \ n_{e, 7} \gamma_3{}^{-1} r_{\rm c, 10}{}^{-1}  r_{\rm emi, 10}{}^2  \ \rm erg \ s^{-1}$. 
By equating the above two, the possible maximum Lorentz factor is given as $\gamma \sim 1.5 \times 10^4 \ n_{e, 7} B_{\rm emi, 5}{}^{-1}  r_{\rm c, 10}{}^{-1} r_{\rm emi, 10} \Omega_0{}^{-1}$, 
which is high enough.      

Comparing the required number density, $n_{\rm e} \sim 10^7 \ \rm cm^{-3}$  with the Goldreich-Julian density $n_{\rm GJ} \approx B_{\rm emi}\Omega/2\pi c e$~\citep{Goldreich_Julian_1969}, 
one obtains the necessary multiplicity as 
\begin{equation}
\kappa_{\rm GJ} = n_{\rm e}/n_{\rm GJ} \sim 8.2 \times 10^3 \ n_{\rm e, 7} B_{\rm emi, 5}{}^{-1} \Omega_0{}^{-1}.
\end{equation}
Although uncertainties in calculating the multiplicity are large, $\kappa_{\rm GJ} \gtrsim 10^4$ is inferred for relatively young NS pulsars~\citep[e.g.][]{Hibschman_Aron_2001}.  
This may also be the case for newly born WD pulsars given that magnetic pair production avalanches can be triggered in the polar region like NS pulsars~\citep{Kashiyama_et_al_2011}.

\section{Summary and discussion}

In summary, the proposed binary WD merger scenario can potentially explain the main observed 
characteristics of FRBs, including the energetics, duration, and event rate.  

In our model, the FRBs may be produced only once, multiple events from the same source are likely inhibited for the following reasons.  
First, the rapidly rotating WD can settle down in a dynamical timescale via, e.g., an angular-momentum transfer to the surrounding debris disk.    
Then the emergence of the inner magnetic field to the polar region would be suppressed, which limits the energy budget for the FRB phenomenon.  
Also, the differential rotation becomes less significant, and the magnetic-field twisting in the polar region, which triggers the energy dissipation via the reconnection, becomes ineffective. 
Finally, the electric potential in the polar region becomes small, and electron bunches with a sufficient multiplicity and Lorentz factor are no longer supplied. 

A key to identifying the FRB progenitor is to detect the counterparts in other wavelengths, which have not been reported so far. 
Based on our scenario, one may expect X-ray counterparts due to fallback accretion of tidally disrupted matter onto the disk surrounding the merged WD.
Numerical simulations set an upper limit for such accretion luminosities as $L_{\rm x} < 10^{47} \ \rm erg \ s^{-1}$ 
with a duration of $\sim 100 \ \rm s$ for an equal mass binary of $\sim 0.6 M_{\odot}$~\citep{Loren_et_al_2009}. 
For a cosmological distance, e.g., $z = 0.5$, the anticipated flux of $\sim 10^{-10} \ \rm erg \ s^{-1} cm^{-2}$ 
is well below the typical trigger threshold of Swift BAT $\sim 10^{-7} \ \rm erg \ s^{-1} \ cm^{-2}$ in $15\mbox{-}150 \ \rm keV$~\citep{Grupe_et_al_2013}. 
However, given the large event rate, one can also expect relatively close events, which are detectable in X rays. 

A SN Ia is another possible counterpart of an FRB in our model. 
In the so-called double degenerate model, a binary WD merger triggers a deflagration and a detonation in some cases, 
resulting in a thermo-nuclear explosion of the merged WD~\citep[see, e.g.,][]{Hillebrandt_Niemeyer_2000}. 
We strongly encourage SN surveys in the field of view of FRB radio observations, although no detection does not exclude our model.  

If SNe Ia and/or the host galaxy were to be observed simultaneously with FRBs, the redshift and the luminosity distance would be determined independently.  
By combining the dispersion measure of the FRBs, one could put unique constraints on cosmological parameters 
(since the luminosity distance and the dispersion measure have different dependences on the redshift) and the cosmological reionization history~\citep{Ioka_2003,Inoeu_2004}. 
To this end, long-baseline FRB observations with a better angular resolution than single dish observations ($\sim (c/{\rm GHz}) \times (64 \ \rm m)^{-1} \gtrsim 0.1 \ \rm deg $) 
are important for finding counterparts and host galaxies.

\acknowledgments
We thank Kohta Murase, Shuta Tanaka, and Shota Kisaka for discussions. 
This work is supported by a JSPS fellowship (KK),  NASA NNX13AH50G (P.M. and K.K.), 
and the Grant-in-Aid for Scientific Research No.24103006, 24000004, 22244030 of Japanese MEXT (K.I.).  


\end{document}